\documentclass[a4paper,11pt]{article}
\usepackage{jheppub} 
\usepackage{lineno}
\usepackage{hyperref}
\usepackage{graphicx}
\usepackage{subcaption}
\usepackage{adjustbox}
\usepackage{xcolor}
\usepackage{placeins}
\usepackage{bm}
\usepackage{tikz}
\usepackage{mathabx}
\usepackage{tikz-feynman}
\usepackage{multirow}
\usepackage{float}
\usepackage{placeins}
\usepackage{subcaption} 
\usepackage{mathtools,slashed}
\usepackage{booktabs}

\tikzset{
    fermion/.style={line width=0.8pt, postaction={decorate}, decoration={markings, mark=at position .55 with {\arrow[scale=1.5]{>}}}},
    noarrowfermion/.style={line width=0.4pt},
    boson/.style={line width=0.8pt, decorate, decoration={snake, segment length=4pt, amplitude=3pt}},
    scalar/.style={dashed, line width=0.8pt},
}
\tikzfeynmanset{compat=1.1.0}

\DeclareMathAlphabet{\mathsfit}{\encodingdefault}{\sfdefault}{m}{sl}

\newcommand{\FDF}[1][]{\varphi^\dagger #1\!\overleftrightarrow{D}\!_\mu\varphi}

\let\Re\undefined

\DeclareMathOperator{\Re}{Re}

\title{\boldmath Sensitivity to top-quark couplings in diboson production at lepton colliders}

\author[a,b]{Eugenia Celada,}
\author[c]{V\'ictor Miralles,}%
\author[b]{Eleni Vryonidou }%

\affiliation[a]{Department of Physics and Astronomy, University of Manchester, Oxford Road, Manchester M13 9PL, United Kingdom}

\affiliation[b]{Department of Physics, University of Cyprus, Panepistimiou Street 1, Aglantzia, CY-1678 Nicosia, Cyprus}

\affiliation[c]{Departament de F\'isica, Universitat d'Alacant,
Campus de Sant Vicent del Raspeig, E-03690 \mbox{Alacant}, Spain}

\emailAdd{eugenia.celada@manchester.ac.uk}
\emailAdd{victor.miralles@ua.es}
\emailAdd{vryonidou.eleni@ucy.ac.cy}

\abstract{We study the next-to-leading order electroweak corrections to $e^+e^-\to W^+W^-$ from dimension-six two-fermion top-quark operators in the Standard Model Effective Field Theory. 
We compute analytical and numerical results for future electron–positron colliders, focusing on the proposed LEP3 and FCC-ee that will operate at centre-of-mass energies below the $t\bar t$ production threshold.
We compare the indirect sensitivity arising from virtual corrections to $WW$ production to that from $ZH$ production, and to the current constraints from LEP and LHC data. We show that NLO corrections can provide competitive sensitivity to these operators. This work represents a first step towards the systematic computation of electroweak corrections to $W$-pair production at lepton colliders in the SMEFT, whose impact can then be properly assessed in global analyses. }

\begin{document}
\maketitle

\section{Introduction}
\label{sec:intro}

The top quark, with a mass of the order of the electroweak scale and a Yukawa coupling of $\mathcal{O}(1)$, is an excellent laboratory to probe new physics (NP) extensions connected to electroweak symmetry breaking.
Any deviation of top-quark couplings from their Standard Model (SM) predictions would provide a clear indication of NP.
Mostly thanks to the LHC programme, many of the possible beyond the SM top-quark couplings have already been constrained~\cite{Buckley:2015nca,Buckley:2015lku,Brivio:2019ius,Bissmann:2019qcd,Hartland:2019bjb,Ellis:2020unq,Ethier:2021bye,Miralles:2021dyw,Alasfar:2022zyr,Aoude:2022aro,Bartocci:2024fmm,Maltoni:2024dpn,Miralles:2024huv,CMS:2024kvw,terHoeve:2025gey, deBlas:2025xhe,Durieux:2019rbz,Celada:2024mcf,Cornet-Gomez:2025jot,Armadillo:2026mvp,Celada:2026ubm,Cornet-Gomez:2026tin}, though the precision remains limited in some cases.

Looking beyond the LHC era, a high-luminosity lepton collider provides a compelling path to precision studies of the Higgs and electroweak sectors.
In particular, the European particle physics community has expressed strong support for the Future Circular Collider programme, starting with an initial $e^+e^-$ phase (FCC-ee)~\cite{deBlas:2025gyz,FCC:2025lpp}.
Runs at and above the $t\bar t$ threshold would enable high-precision determinations of top-quark couplings~\cite{Durieux:2018tev,Durieux:2019rbz,deBlas:2022ofj,Celada:2024mcf,Armadillo:2026mvp,Cornet-Gomez:2025jot}.
However, even below the $t\bar t$ threshold, the ``descoped" FCC-ee would retain indirect sensitivity to the top-quark sector through loop effects in precision processes.
Moreover, a cost-effective alternative is an $e^+e^-$ machine in the existing LHC tunnel (the so-called LEP3 concept), whose centre-of-mass energy would not reach the $t\bar t$ threshold~\cite{Anastopoulos:2025jyh}.
This strengthens the case for exploiting indirect probes of top-quark interactions at lepton colliders operating as Higgs/electroweak factories.

Indirect sensitivity to top-quark couplings below the $t\bar t$ threshold arises through Electroweak Precision Observables (EWPOs), Higgs production and decay, as well as $W$-pair production. Such sensitivity to top-quark couplings is typically quantified by employing the Standard Model Effective Field Theory (SMEFT) framework, which allows us to parametrise deviations from the SM predictions in a model-independent manner. The need for a careful assessment of indirect one-loop sensitivity has motivated a campaign of NLO computations in the SMEFT. Corrections of electroweak (EW) origin, unlike NLO QCD corrections, are not yet available in an automated framework for generic processes and computations have been performed on a process-by-process basis. 
 
The complete SMEFT NLO contributions have been computed for Higgs decays ~\cite{Hartmann:2015oia,Ghezzi:2015vva,Hartmann:2015aia,Gauld:2015lmb,Gauld:2016kuu,Dawson:2018pyl,Dedes:2018seb,Dawson:2018liq,Dawson:2018jlg,Dawson:2024pft,Bellafronte:2025jbk}
and for the Electroweak Precision Observables~\cite{Dawson:2019clf,Dawson:2022bxd,Bellafronte:2023amz,Biekotter:2025nln}.
Also, a code providing a full implementation of the electroweak corrections to Higgs decays has become recently available~\cite{Bellafronte:2026mhp}. Additionally the Higgsstrahlung process $e^+e^-\to ZH$  has been computed at NLO  \cite{Asteriadis:2024xuk,Asteriadis:2024xts}. 
These studies suggest that the well-known logarithmic running of the
Wilson coefficients \cite{Jenkins:2013zja,Jenkins:2013wua,Alonso:2013hga} captures only part of the full one-loop SMEFT corrections. The finite terms can modify the normalisation and kinematic dependence of the observables, generate sensitivity to
operators that do not contribute at tree level, and change the correlations in the EFT parameter space. Therefore, relying only on RGE-improved LO predictions may miss phenomenologically
relevant effects, especially at future lepton colliders where the expected experimental precision is comparable to the size of electroweak NLO corrections. As such, the impact of full higher order corrections has to be assessed carefully for each process. 

In the effort of establishing indirect sensitivity to top-quark couplings, $W$-pair production can play a key role.  Within SMEFT, at tree level, this process is mostly sensitive to bosonic and light-quark two-fermion operators. Its tree-level impact in the context of global fits has been studied at future colliders both inclusively and fully differentially~\cite{deBlas:2022ofj,DeBlas:2019qco,Celada:2024mcf}.
$W$-pair production will be probed with high accuracy at future $e^+e^-$ colliders above the production threshold ($\sqrt{s}=161$ GeV) and can offer an unprecedented sensitivity to higher-order effects as suggested also in previous SMEFT studies~\cite{Vryonidou:2018eyv,Durieux:2018ggn}. This motivates a systematic computation of this process at one-loop in the presence of top-quark operators which  constitutes the main goal of this work. 

In this paper we study the contributions of top quark two-fermion operators in the SMEFT to diboson production at prospective future lepton colliders. In particular, we explicitly compute the NLO EW corrections to $e^+e^-\to W^+W^-$ induced by top-quark two-fermion operators, as well as $e^+e^-\to ZH$, which was previously computed in Ref.~\cite{Asteriadis:2024xts}. 
In Section~\ref{sec:theoframework} we review the theoretical setup and the details of the NLO renormalisation. We then present our numerical results in Section~\ref{sec:results}, and we compare the projected bounds at future colliders with the current constraints. We finally draw our conclusions and outlook in Section~\ref{sec:conclusions}.

\section{Theoretical Framework}
\label{sec:theoframework}
In the SMEFT, deviations from the SM are parametrised by effective operators of dimension $d>4$, organised in a series expansion of a high energy scale $\Lambda$
\begin{equation}
    \mathcal{L}_{\rm{SMEFT}} = \mathcal{L}_{\rm{SM}} +  \sum_{i} \frac{C_i O_i}{\Lambda^{2}} + \mathcal{O}(\Lambda^{-4}) \, .
\end{equation}
The first relevant contributions are those at dimension six. In the Warsaw basis~\cite{Grzadkowski:2010es}, the relevant deviations of top-quark couplings are generated by
\begin{alignat}{3}
    \nonumber
    O_{\varphi q}^{(1)\,[33]}
		&  \equiv 
		\left(\FDF[i] \right) \left(\bar q_3 \gamma^\mu q_3 \right)	,\qquad
	&O_{\varphi q}^{(3)\,[33]}
		&\equiv \left( \varphi^\dagger \tau_I  i\!\overleftrightarrow{D}\!_\mu \varphi \right) (\bar q_3  \gamma^\mu \tau^I q_3) ,\\
    \label{eq:op_2q}
    O_{\varphi u}^{[33]}
		&\equiv  
		\left(\FDF[i] \right) \left(\bar u_3 \gamma^\mu u_3 \right)	,
    &O_{\varphi ud}^{[33] \, (*)}
        &\equiv \left(\widetilde{\varphi}^{\dagger} i D_\mu \varphi \right)
        	\left(\bar u_3 \gamma^\mu d_3 \right),\\
    \nonumber
    O_{uW}^{[33] \, (*)}
		&\equiv 
		\left(\bar q_3 \sigma^{\mu\nu} u_3  \right) \tau_I \widetilde{\varphi} \, W_{\mu\nu}^I ,
    &O_{uB}^{[33] \, (*)}
		&\equiv 
		\left(\bar q_3 \sigma^{\mu\nu} u_3  \right) \widetilde{\varphi} \, B_{\mu\nu} ,\\
    \nonumber
    O_{u\varphi}^{[33] \, (*)}
		&\equiv 
		\left( \varphi^\dagger \varphi \right) \bar q_3 \, u_3 \,\widetilde{\varphi}  ,
    & &
\end{alignat}
where we write explicitly the flavour indices. The operators marked with $^{(*)}$ are not hermitian and have a complex coefficient.
However, we consider all Wilson coefficients to be real, and we drop the $^{(*)}$ from now on to simplify the notation. 
In our convention the covariant derivative is
\begin{equation}
    D_{\mu} = \partial_{\mu} +i g \tau^I W^I_{\mu} +i g' Y B_{\mu}
\end{equation}
where $g'$ and $g$ denote the $U(1)_Y$ and $SU(2)_L$ gauge couplings respectively, and 
\begin{align}
    \FDF[i] &= \varphi^\dagger i(D_{\mu} \varphi) - i(D_{\mu} \varphi^\dagger)\varphi \, , \\
     \varphi^\dagger \tau_I i\!\overleftrightarrow{D}\!_\mu \varphi &= \varphi^\dagger \tau_I i(D_{\mu} \varphi) - i(D_{\mu} \varphi^\dagger) \tau_I\varphi \, ,
\end{align}
while $\widetilde{\varphi} = 2i \tau^2 \varphi^*$ is the conjugate Higgs doublet.

Whilst we will present results for the Warsaw coefficients, it is straightforward to extract the corresponding results for the \texttt{dim6top} and \texttt{SMEFT@NLO} bases \cite{Aguilar-Saavedra:2018ksv,Degrande:2020evl}.

\subsection{Tree level operators and input scheme}
We use as input parameters the fermion masses of the heavy quarks, the boson masses and the Fermi constant:
$$
\{m_W,\, m_Z,\, m_H,\, m_b,\, m_t,\, G_\mu\}
$$
while all the leptons and light quarks are considered massless.
A comprehensive comparison of different input choices has been performed for instance in Ref.~\cite{Biekotter:2023xle}.
This choice is commonly referred to as the $\alpha_\mu$ scheme, since the electroweak coupling is derived from the Fermi constant $G_\mu$ measured in muon decay, and is also sometimes denoted as the $m_W$ scheme in the SMEFT literature.

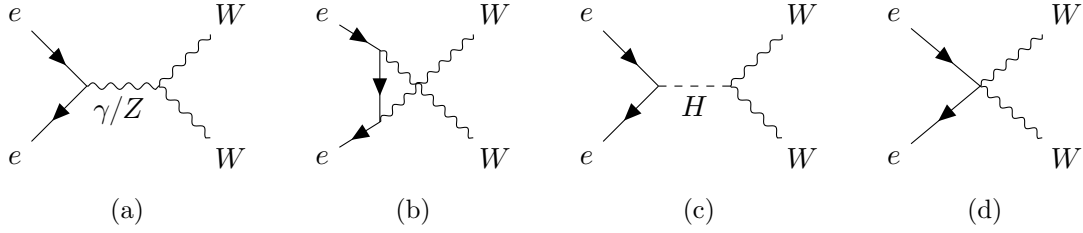
\begin{figure}[t!]
\centering

\begin{subfigure}{0.25\textwidth}
\centering
\begin{tikzpicture}[scale=0.95]
\begin{feynman}
  \vertex (ep) at (-2.0,  1.0) {\(e\)};
  \vertex (em) at (-2.0, -1.0) {\(e\)};
  \vertex (v0) at (-1.0, 0.0) ;  
  \vertex (v2) at ( 0, 0.0) ;   
  \vertex (W1) at ( 1.,  1.0) {\(W\)};
  \vertex (W2) at ( 1., -1.0) {\(W\)};

  \diagram*{
    (ep) -- [fermion] (v0) -- [fermion] (em),
    (v0) -- [photon, edge label'=\(\hspace*{-1 mm}\gamma/Z\)] (v2),
    (v2) -- [boson, edge label=\(\)] (W1),
    (v2) -- [boson, edge label'=\(\)] (W2),
  };
\end{feynman}
\end{tikzpicture}
\caption{}
\end{subfigure}\hfill
\begin{subfigure}{0.25\textwidth}
\centering
\begin{tikzpicture}[scale=0.95]
\begin{feynman}
  \vertex (ep) at (-1.2,  1.0) {\(e\)};
  \vertex (em) at (-1.2, -1.0) {\(e\)};
  \vertex (v0) at (-0.4, 0.5) ;
  \vertex (v2) at (-0.4, -0.5) ;    
  \vertex (W1) at ( 1.2,  1.0) {\(W\)};
  \vertex (W2) at ( 1.2, -1.0) {\(W\)};

  \diagram*{
    (ep) -- [fermion] (v0) -- [fermion] (v2) -- [fermion] (em),
    (v2) -- [boson, edge label=\(\)] (W1),
    (v0) -- [boson, edge label'=\(\)] (W2),
  };
\end{feynman}
\end{tikzpicture}
\caption{}
\end{subfigure}\hfill
\begin{subfigure}{0.25\textwidth}
\centering
\begin{tikzpicture}[scale=0.95]
\begin{feynman}
  \vertex (ep) at (-2.0,  1.0) {\(e\)};
  \vertex (em) at (-2.0, -1.0) {\(e\)};
  \vertex (v0) at (-1.0, 0.0) ;
  \vertex (v2) at ( 0, 0.0) ;    
  \vertex (W1) at ( 1.,  1.0) {\(W\)};
  \vertex (W2) at ( 1., -1.0) {\(W\)};

  \diagram*{
    (ep) -- [fermion] (v0) -- [fermion] (em),
    (v0) -- [scalar, edge label'=\(H\)] (v2),
    (v2) -- [boson, edge label=\(\)] (W1),
    (v2) -- [boson, edge label'=\(\)] (W2),
  };
\end{feynman}
\end{tikzpicture}
\caption{}
\end{subfigure}\hfill
\begin{subfigure}{0.25\textwidth}
\centering
\begin{tikzpicture}[scale=0.95]
\begin{feynman}
  \vertex (ep) at (-1.2,  1.0) {\(e\)};
  \vertex (em) at (-1.2, -1.0) {\(e\)};
  \vertex (v0) at (0.0, 0.0) ;  
  \vertex (W1) at ( 1.2,  1.0) {\(W\)};
  \vertex (W2) at ( 1.2, -1.0) {\(W\)};

  \diagram*{
    (ep) -- [fermion] (v0) -- [fermion] (em),
    (v0) -- [boson, edge label=\(\)] (W1),
    (v0) -- [boson, edge label'=\(\)] (W2),
  };
\end{feynman}
\end{tikzpicture}
\caption{}
\end{subfigure}
\caption{Tree-level diagrams for $e^+e^-\to W^+W^-$. Only diagrams (a) and (b) are present in the SM, when the electron is massless. 
}
\label{fig:tree_diagrams_eeWW}
\end{figure}

At tree level, in the $\alpha_\mu$ scheme, the $e^+e^-\rightarrow W^+ W^-$ and $e^+e^-\rightarrow ZH$ processes are sensitive to the following operators in the Warsaw basis:
\begin{alignat}{3}
   \nonumber
   O_{\varphi \Box}
		&\equiv \left(\varphi^{\dagger} \varphi \right)\Box \left(\varphi^{\dagger} \varphi \right) , \quad
	&O_{\varphi D}
		&\equiv \left(\varphi^{\dagger} D^{\mu} \varphi \right)^{\dagger} \left(\varphi^{\dagger} D_{\mu} \varphi \right) ,\\
    \nonumber
    O_{\varphi B}
		&\equiv \left(\varphi^{\dagger} \varphi \right) B^{\mu \nu} B_{\mu \nu} , \quad
	&O_{\varphi W}
		&\equiv \left(\varphi^{\dagger} \varphi \right) W^{\mu \nu}_I W_{\mu \nu}^I ,\\
    \nonumber
    O_{\varphi WB}
		&\equiv \left(\varphi^{\dagger} \tau^I \varphi \right) B^{\mu \nu} W_{\mu \nu}^I , \quad
    &O_{W}
		&\equiv \varepsilon^{IJK} W_\mu^{I\nu} W_\nu^{J\rho} W_\rho^{K\mu} ,\\
    \label{eq:op_tree}
    O_{e W}^{[11]}
		&\equiv (\bar \ell_1 \sigma^{\mu\nu} e_1) \tau^I \varphi W_{\mu\nu}^I , \quad
    &O_{e B}^{[11]}
		&\equiv (\bar \ell_1 \sigma^{\mu\nu} e_1) \varphi B_{\mu\nu} ,\\
    \nonumber
    O_{e \varphi }^{[11]}
		&\equiv (\varphi^\dagger \varphi)(\bar \ell_1 \varphi e_1 ) , \quad
    &O_{\varphi e}^{[11]}
		&\equiv \left(\FDF[i] \right) \left( \bar e_1 \gamma^\mu e_1 \right) ,\\
    \nonumber
    O_{\varphi \ell}^{(3)\,[11]}
		&\equiv \left(  \varphi^\dagger \tau_I i\!\overleftrightarrow{D}\!_\mu \varphi \right) 
        \left(\bar \ell_1  \gamma^\mu \tau^I \ell_1\right) , \quad
   &O_{\varphi \ell}^{(3)\,[22]}
		&\equiv \left(  \varphi^\dagger \tau_I i\!\overleftrightarrow{D}\!_\mu \varphi \right) 
        \left(\bar \ell_2  \gamma^\mu \tau^I \ell_2\right) ,\\
    \nonumber
    O_{\varphi \ell}^{(1)\,[11]}
		&\equiv \left(\FDF[i] \right) \left(\bar \ell_1  \gamma^\mu \ell_1 \right) , \quad
    &O_{\ell \ell}^{[1221]}
		&\equiv \left( \bar \ell_1 \gamma_{\mu} \ell_2 \right)  
        \left( \bar \ell_2 \gamma^{\mu} \ell_1  \right) .
\end{alignat}
These operators, although not the main focus of this study, are relevant in the renormalisation procedure of the NLO amplitude, as some of them mix with top-quark operators due to the renormalisation group evolution (RGE), see Tab.~\ref{tab:top_operator_mixing_matrix}.
They generate tree-level diagrams such as those shown in Fig.~\ref{fig:tree_diagrams_eeWW}, and induce input scheme-dependent shifts in the relations between the chosen input observables and the Lagrangian parameters. 

\subsection{SMEFT renormalisation}

\begin{figure}[t!]
\centering

\begin{subfigure}{0.32\textwidth}
\centering
\begin{tikzpicture}[scale=0.95]
\begin{feynman}
  \vertex (ep) at (-2.0,  1.0) {\(e\)};
  \vertex (em) at (-2.0, -1.0) {\(e\)};
  \vertex (v0) at (-1.2, 0.0) ;
  \vertex[dot] (v1) at ( 0.0, 0.0) {};   
  \vertex (v2) at ( 1.2, 0.0) ;   
  \vertex (v3) at ( 0.0, 0.0) ;   
  \vertex (W1) at ( 2.,  1.0) {\(W\)};
  \vertex (W2) at ( 2., -1.0) {\(W\)};

  \diagram*{
    (ep) -- [fermion] (v0) -- [fermion] (em),

    (v0) -- [photon, edge label'=\(\hspace*{-1 mm}\gamma/Z\)] (v1),

    (v1) -- [fermion, out=50, in=130, looseness=50.,
             edge node={node[midway, above] {\(t\)}}] (v3),

    (v1) -- [scalar, edge label'=\(H\)] (v2),

    (v2) -- [boson, edge label=\(\)] (W1),
    (v2) -- [boson, edge label'=\(\)] (W2),
  };
\end{feynman}
\end{tikzpicture}
\caption{}
\end{subfigure}\hfill
\begin{subfigure}{0.32\textwidth}
\centering
\begin{tikzpicture}[scale=0.95]
\begin{feynman}
  \vertex (e1) at (-2.0,  1.0) {\(e\)};
  \vertex (e2) at (-2.0, -1.0) {\(e\)};
  \vertex (v0) at (-1.2,  0.0);
  \vertex[dot] (v1) at (-0.2,  0.0) {};
  \vertex[crossed dot, minimum size=5pt, inner sep=0pt] (v2) at ( 0.9,  0.0) {};
  \vertex (v3) at ( 1.7,  0.0);
  \vertex (W)  at ( 2.5,  1.0) {\(W\)};
  \vertex (nu) at ( 2.5, -1.0) {\(W\)};

  \diagram*{
    (e1) -- [fermion] (v0) -- [fermion] (e2),
    (v0) -- [photon, edge label=\(\hspace*{-1 mm}\gamma/Z\)] (v1),

    (v1) -- [fermion, half left, looseness=1.05, edge label=\(t\)] (v2),
    (v2) -- [fermion, half left, looseness=1.05, edge label'=\(t\)] (v1),

    (v2) -- [boson, edge label=\(\)] (v3),
    (v3) -- [boson, edge label=\(\)] (W),
    (v3) -- [boson, edge label'=\(\)] (nu),
  };
\end{feynman}
\end{tikzpicture}
\caption{}
\end{subfigure}\hfill
\begin{subfigure}{0.32\textwidth}
\centering
\begin{tikzpicture}[scale=0.95]
\begin{feynman}
  \vertex (e1) at (-2.0,  1.0) {\(e\)};
  \vertex (e2) at (-2.0, -1.0) {\(e\)};
  \vertex (v0) at (-1.2,  0.0);
  \vertex (v1)[dot] at (-0.2,  0.0) {};
  \vertex (v2)[crossed dot, minimum size=5pt, inner sep=0pt] at ( 1.0,  0.0) {};
  \vertex (W1) at ( 2.2,  1.0) {\(W\)};
  \vertex (W2) at ( 2.2, -1.0) {\(W\)};

  \diagram*{
    (e1) -- [fermion] (v0) -- [fermion] (e2),
    (v0) -- [photon, edge label=\(\hspace*{-1 mm}\gamma/Z\)] (v1),

    (v1) -- [fermion, half left, looseness=1.05, edge label=\(t\)] (v2),
    (v2) -- [fermion, half left, looseness=1.05, edge label'=\(t\)] (v1),

    (v2) -- [boson, edge label=\(W\)] (W1),
    (v2) -- [boson, edge label'=\(W\)] (W2),
  };
\end{feynman}
\end{tikzpicture}
\caption{}
\end{subfigure}

\vspace{2mm}

\begin{subfigure}{0.32\textwidth}
\centering
\begin{tikzpicture}[scale=0.95]
\begin{feynman}
  \vertex (e1) at (-2.0,  1.0) {\(e\)};
  \vertex (e2) at (-2.0, -1.0) {\(e\)};
  \vertex (v0) at (-1.2,  0.0);
  \vertex(v1) at (-0.2,  0.0) ;
  \vertex[crossed dot, minimum size=5pt, inner sep=0pt]  (u)  at ( 0.6,  0.9) {};
  \vertex[dot] (d)  at ( 0.2, -0.1) {};
  \vertex (W1) at ( 1.8,  1.0) {\(W\)};
  \vertex (W2) at ( 1.8, -1.0) {\(W\)};

  \diagram*{
    (e1) -- [fermion] (v0) -- [fermion] (e2),
    (v0) -- [photon, edge label=\(\hspace*{-5 mm}\gamma/Z\)] (d),

    (u)  -- [fermion, half left, looseness=1.05] (d),
    (d)  -- [fermion, half left, looseness=1.05, edge label'=\(\hspace*{-1 mm}b\hspace*{6 mm} t\)] (u),

    (u) -- [boson, edge label=\(\)] (W1),
    (d) -- [boson, edge label'=\(\)] (W2),
  };
\end{feynman}
\end{tikzpicture}
\caption{}
\end{subfigure}\hfill
\begin{subfigure}{0.32\textwidth}
\centering
\begin{tikzpicture}[scale=0.95]
\begin{feynman}
  \vertex (e1) at (-2.0,  1.0) {\(e\)};
  \vertex (e2) at (-2.0, -1.0) {\(e\)};
  \vertex (v0) at (-1.2,  0.0);
  \vertex[dot] (v1) at (-0.2,  0.0) {};
  \vertex[crossed dot, minimum size=5pt, inner sep=0pt] (v2) at ( 0.9,  0.0) {};
  \vertex (v3) at ( 1.7,  0.0);
  \vertex (W1) at ( 2.5,  1.0) {\(W\)};
  \vertex (W2) at ( 2.5, -1.0) {\(W\)};

  \diagram*{
    (e1) -- [fermion] (v0) -- [fermion] (e2),
    (v0) -- [photon, edge label=\(\hspace*{-1 mm}\gamma/Z\)] (v1),

    (v1) -- [fermion, half left, looseness=1., edge label=\(t\)] (v2),
    (v2) -- [fermion, half left, looseness=1., edge label'=\(t\)] (v1),

    (v2) -- [scalar, edge label=\(H\)] (v3),
    (v3) -- [boson, edge label=\(\)] (W1),
    (v3) -- [boson, edge label'=\(\)] (W2),
  };
\end{feynman}
\end{tikzpicture}
\caption{}
\end{subfigure}\hfill
\begin{subfigure}{0.32\textwidth}
\centering
\begin{tikzpicture}[scale=0.95]
\begin{feynman}
  \vertex (e1) at (-2.0,  1.0) {\(e\)};
  \vertex (e2) at (-2.0, -1.0) {\(e\)};
  \vertex (v0) at (-1.2,  0.0);
  \vertex[dot] (v1) at (-0.2,  0.0) {};

  \vertex[crossed dot, minimum size=5pt, inner sep=0pt] (u)  at ( 0.9,  0.7){};
  \vertex (d)  at ( 0.9, -0.7);

  \vertex (W1) at ( 2.2,  1.0) {\(W\)};
  \vertex (W2) at ( 2.2, -1.0) {\(W\)};

  \diagram*{
    (e1) -- [fermion] (v0) -- [fermion] (e2),
    (v0) -- [photon, edge label=\(\hspace*{-1 mm}\gamma/Z\)] (v1),

    (v1) -- [fermion, edge label=\(t\)] (u),
    (v1) -- [anti fermion, edge label'=\(t\)] (d),

    (u) -- [fermion, edge label=\(b\)] (d),

    (u) -- [boson, edge label=\(\)] (W1),
    (d) -- [boson, edge label'=\(\)] (W2),
  };
\end{feynman}
\end{tikzpicture}
\caption{}
\end{subfigure}

\caption{Representative one-loop contributions of dimension six top-quark operators to $e^+e^-\to W^+W^-$. The marked vertices represent possible insertions of dimension six operators. Only one insertion is considered at a time. }
\label{fig:top_diagrams_eeWW}
\end{figure}
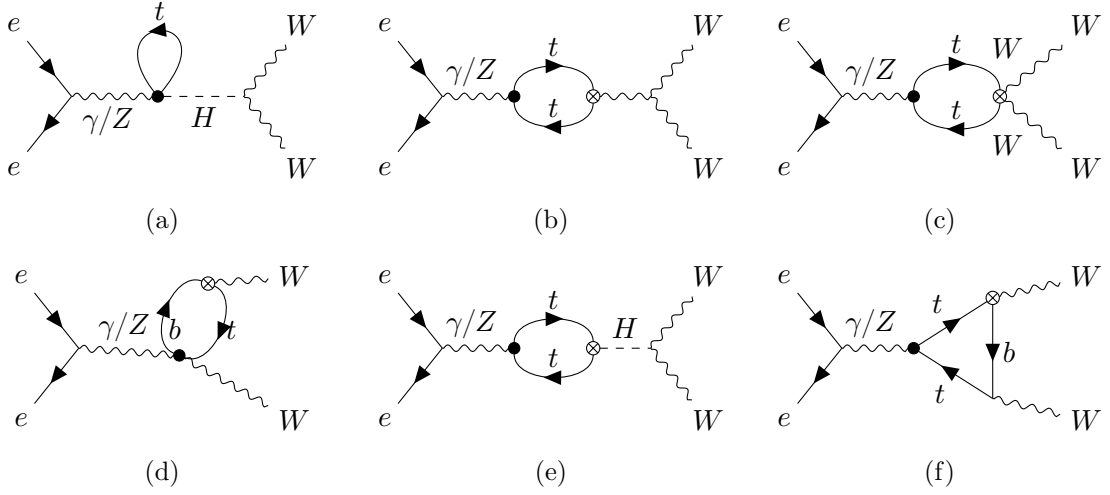

In this section we describe the computation of the one loop $W^+ W^-$ production amplitude.
Retaining only up to linear SMEFT contributions, the total NLO amplitude is given by
\begin{equation}
\label{eq:amplitude}
    \mathcal{M}_{\rm NLO} =  \mathcal{M}^{(4)}_{\rm LO} + \sum_i \frac{C_i(\mu)}{\Lambda^2} \mathcal{M}^{i,(6)}_{\rm LO} + \mathcal{M}^{(4)}_{\rm NLO} + \sum_j \frac{C_j(\mu)}{\Lambda^2} \mathcal{M}^{j,(6)}_{\rm NLO} \, ,
\end{equation}
where the superscripts (4) and (6) denote the SM and SMEFT contributions, and the indices $i$ and $j$ run respectively over the tree level coefficients in Eq.~\ref{eq:op_tree} and the top-quark ones of Eq.~\ref{eq:op_2q}.
The NLO calculation requires two parts: the bare one-loop amplitude and the counterterm to reabsorb the UV divergences, so that the renormalised amplitude for a top-quark operator $O_i$ is given by
\begin{equation}
    \mathcal{M}^{i,(6)}_{\rm NLO} = \mathcal{M}_V^{i,(6)} + \delta \mathcal{M}_{\rm SM}^{i,(6)} + \delta \mathcal{M}_{\rm RGE}^{i,(6)} .
\end{equation}
The virtual contribution $\mathcal{M}_V$ is the UV divergent one loop amplitude from dimension-six operators. The SM counterterm $\delta \mathcal{M}_{\rm SM}^{(6)}$ renormalises the SM couplings and wavefunctions at dimension-six, while $\delta \mathcal{M}_{\rm RGE}^{(6)}$ is the counterterm that renormalises Wilson coefficients according to the anomalous dimension matrix.
We use a mixed renormalisation scheme where SM parameters are renormalised on-shell (OS) \cite{Denner:1991kt} while Wilson coefficients are renormalised in the modified minimal subtraction scheme ($\overline{\mathrm{MS}}$). 
\paragraph{Virtual amplitude.}
The operators of Eq.~\ref{eq:op_2q} enter the $e^+ e^- \to W^+ W^-$ amplitude at NLO and generate heavy fermion loops like those shown in Fig.~\ref{fig:top_diagrams_eeWW}. The current operators $O_{\varphi q}^{(1)\,[33]}$ and $O_{\varphi q}^{(3)\,[33]}$ generate top and bottom loops in the mass basis, and both fermion loops 
are needed to renormalise these operators.
We generated these diagrams using the \texttt{FeynArts} \cite{Hahn:2000kx} model generated with the \texttt{FeynRules} \cite{Alloul:2013bka} package \texttt{SmeftFR} \cite{Dedes:2023zws}. The analytical computation is performed with \texttt{FeynCalc} \cite{Mertig:1990an,Shtabovenko:2016sxi,Shtabovenko:2020gxv,Shtabovenko:2023idz} and the evaluation of the Passarino-Veltman functions is then carried out with \texttt{Package-X}~\cite{Patel:2016fam} and \texttt{LoopTools} \cite{Hahn:1998yk}, verifying that we get the same numerical results.

Special care is required in the treatment of Dirac traces involving $\gamma_5$.
The matrix $\gamma_5 = \frac{i}{4!}\,\varepsilon_{\mu\nu\rho\sigma}\gamma^\mu\gamma^\nu\gamma^\rho\gamma^\sigma$ is intrinsically four-dimensional, since the Levi-Civita tensor $\varepsilon_{\mu\nu\rho\sigma}$ is defined only in four dimensions.
Its presence leads to well-known consistency issues in dimensional regularisation, for which different prescriptions have been proposed in the literature.
In the Kreimer, Körner and Schilcher (KKS) scheme \cite{Chanowitz:1979zu,Kreimer:1989ke,Korner:1991sx}, sometimes referred to as naive dimensional regularisation (NDR), the proposed solution is to preserve the anti-commutation between the $\gamma_5$ matrix and the other $\gamma^{\mu}$ matrices in $d$-dimensions. With this definition, the cyclicity of the trace is lost. An alternative scheme, proposed by Breitenlohner, Maison, ’t-Hooft and Veltman (BMHV) \cite{tHooft:1972tcz,Breitenlohner:1977hr}, preserves the cyclicity of the trace at the price of a non-vanishing anticommutator $\{\gamma_5, \gamma^{\mu}\} \neq 0$ in $d$-dimensions. 
In this work we use the KKS scheme, meaning that traces with an odd number of $\gamma_5$ matrices and at least six Dirac matrices are not well defined, and their evaluation necessitates a reading point prescription. Since cyclicity is lost, the result will depend on the choice of the reading point, which states the first $\gamma^\mu$ matrix of the trace. While this choice is not physical and in principle arbitrary, different reading point prescriptions generate differences of order $\mathcal{O}(\epsilon)$ in the evaluation of the trace that translate into a different finite term for UV-divergent amplitudes. As long as the same prescription is assumed, these differences are cancelled after performing the matching to renormalisable UV models. We therefore choose to follow the same prescription used in the matching code \texttt{Matchete} \cite{Fuentes-Martin:2020udw,Fuentes-Martin:2022jrf}, which is to use as a reading point the vertex of the Wilson coefficient. This guarantees the consistency and applicability of our results, so that our bounds can be easily translated to UV-complete models by performing the matching with the same prescription. 
\paragraph{SM counterterm.}
The SM counterterm  $\delta \mathcal{M}_{\rm SM}^{(6)}$ is obtained by renormalising the SM fields and independent parameters that appear in the tree level SM amplitude~\cite{Denner:1991kt}. 
The bare masses, couplings and external fields, denoted as $X_{0}$, are related to the renormalised ones by renormalisation constants $\delta X$ according to
\begin{align}
    m_{V,0}^2 &= m_{V}^2+ \delta m_{V}^2 \, , \qquad V=W,Z \\
    G_{\mu,0} &= G_{\mu} + \delta G_{\mu} \, , 
\end{align}
for the input parameters and
\begin{align}
    &W^{\pm}_0 =  \left(1 + \frac{1}{2} \delta Z_W \right)W^{\pm} \, , \\
    &e_{L/R,0}^{\pm} = \left( 1 + \frac{1}{2} \delta Z_e^{L/R} \right) e_{L/R}^{\pm} \, 
\end{align}
for the external fields. We notice that because we assume the electron to be massless we do not have to renormalise its mass. Moreover, the fermionic wavefunction $e_{L/R,0}^{\pm}$ at one loop does not receive contributions by any of the top-quark operators considered and therefore does not enter the renormalisation procedure.
The renormalisation constants are defined by computing the two-point functions 1PI contributions, $\Sigma$. 
The one loop self-energies have a SM and a dimension-6 contribution from top-quark operators, $\Sigma = \Sigma^{(4)}+\Sigma^{(6)}$.
In the 't Hooft-Feynman gauge the NLO two-point functions of interest are 
\begin{align}
    \Gamma^W_{\mu \nu}(k) &= -i g_{\mu \nu} (k^2-m_W^2)-i \left( g_{\mu \nu} - \frac{k_{\mu}k_{\nu}}{k^2} \right)\Sigma^W_T(k^2) -i \frac{k_{\mu}k_{\nu}}{k^2} \Sigma^W_L(k^2) \, , \\
    \Gamma^{Z}_{\mu \nu}(k) &= -i g_{\mu \nu} (k^2-m_Z^2)-i \left( g_{\mu \nu} - \frac{k_{\mu}k_{\nu}}{k^2} \right)\Sigma^{Z}_T(k^2) -i \frac{k_{\mu}k_{\nu}}{k^2} \Sigma^{Z}_L(k^2) \, .
\end{align}
 The OS conditions are imposed on the renormalised two-point functions and determine the mass and field-renormalisation constants. For instance, for the electroweak gauge bosons we use
\begin{align}
    \delta m_V^2 &= \Re \Sigma_T^{VV}(m_V^2) \, , \qquad V=W,Z \, ,\\
    \delta Z_W &= -\Re \frac{\partial \Sigma_T^{WW}(k^2)}
    {\partial k^2} \bigg|_{k^2=m_W^2} \, .
\end{align}

The renormalisation constant of $G_\mu$ is fixed from muon decay by requiring that, at zero momentum transfer, the radiatively corrected amplitude is normalised as the tree-level Fermi interaction. This condition determines $\delta G_\mu$.

\paragraph{Wilson coefficient counterterms.}
The SMEFT Wilson coefficients are renormalised in the $\overline{\mathrm{MS}}$ scheme, according to 
\begin{equation}
\label{eq:RGE}
    C_i(\mu) = C_{i,0}- \delta C_i = 
    C_{i,0} -\frac{1}{2 \epsilon} \left(\frac{\mu^2 }{4\pi}e^{\gamma_E}\right)^{\epsilon} \frac{1}{16\pi^2}\gamma_{ij} C_j(\mu) \, ,
\end{equation}
where the running of the coupling is governed by the anomalous dimension matrix coefficients $\gamma_{ij}$, as defined in the
one loop RGE~\cite{Jenkins:2013zja,Jenkins:2013wua,Alonso:2013hga}
\begin{equation}
    \dot{C_i}(\mu) = \frac{d C_i (\mu)}{d \log\left(\mu\right)} = \frac{1}{16\pi^2}\gamma_{ij} C_j(\mu) \, .
\end{equation}
At leading order, operator mixing induces additional contributions through the renormalisation of the Wilson coefficients, which must be taken into account. Since the tree-level amplitude depends on the bare Wilson coefficients of Eq.~\ref{eq:op_tree}, expressing these bare parameters in terms of the renormalised coefficients generates counterterm contributions proportional to the anomalous dimension matrix, and hence to the top-quark operators that mix into them. 
In our computations, the relevant contributions of the anomalous dimension are collected in Tab.~\ref{tab:top_operator_mixing_matrix}.
\begin{table}[t]
\centering
\renewcommand{\arraystretch}{1.6}
\setlength{\tabcolsep}{4pt}
\resizebox{\textwidth}{!}{%
\begin{tabular}{|c|ccccccc|}
\toprule
&
$C_{\varphi q}^{(1)[33]}$
&
$C_{\varphi q}^{(3)[33]}$
&
$C_{\varphi u}^{[33]}$
&
$C_{\varphi ud}^{[33]}$
&
$C_{uB}^{[33]}$
&
$C_{uW}^{[33]}$
&
$C_{u\varphi}^{[33]}$
\\
\midrule

$C_{\varphi WB}$
&
$0$
&
$0$
&
$0$
&
$0$
&
$6g y_t$
&
$10g' y_t$
&
$0$
\\

$C_{\varphi D}$
&
$\displaystyle \frac{8}{3}g'^2 + 24y_t^2 - 24y_b^2$
&
$0$
&
$\displaystyle \frac{16}{3}g'^2 - 24y_t^2$
&
$-24y_t y_b$
&
$0$
&
$0$
&
$0$
\\

$C_{\varphi \Box}$
&
$\displaystyle \frac{2}{3}g'^2 + 6y_t^2 - 6y_b^2$
&
$\displaystyle 6g^2 -18y_t^2 -18y_b^2$
&
$\displaystyle \frac{4}{3}g'^2 - 6y_t^2$
&
$12y_t y_b$
&
$0$
&
$0$
&
$0$
\\

$C_{\varphi B}$
&
$0$
&
$0$
&
$0$
&
$0$
&
$-10g' y_t$
&
$0$
&
$0$
\\

$C_{\varphi W}$
&
$0$
&
$0$
&
$0$
&
$0$
&
$0$
&
$-6g y_t$
&
$0$
\\

$C_{\varphi e}^{[11]}$
&
$\displaystyle -\frac{4}{3}g'^2$
&
$0$
&
$\displaystyle -\frac{8}{3}g'^2$
&
$0$
&
$0$
&
$0$
&
$0$
\\

$C_{\varphi l}^{(1)[11]}$
&
$\displaystyle -\frac{2}{3}g'^2$
&
$0$
&
$\displaystyle -\frac{4}{3}g'^2$
&
$0$
&
$0$
&
$0$
&
$0$
\\

$C_{\varphi l}^{(3)[11]}$
&
$0$
&
$2g^2$
&
$0$
&
$0$
&
$0$
&
$0$
&
$0$
\\
$C_{\varphi l}^{(3)[22]}$
&
$0$
&
$2g^2$
&
$0$
&
$0$
&
$0$
&
$0$
&
$0$
\\
\bottomrule
\end{tabular}%
}
\caption{Anomalous dimension coefficients $\gamma_{ij}$, as given in~\cite{Jenkins:2013zja,Jenkins:2013wua,Alonso:2013hga}, relevant for the renormalisation of operators in Eq.~\ref{eq:op_2q}.}
\label{tab:top_operator_mixing_matrix}
\end{table}

\section{Results}
\label{sec:results}

In this work, we focus on the linear contributions in the EFT expansion, $\mathcal{O}(1/\Lambda^{2})$, given by the interference of the UV-renormalised NLO amplitude at dimension six $\mathcal{M}^{i,(6)}_{\rm NLO}$ and the tree level SM amplitude $\mathcal{M}_{\rm LO}^{(4)}$ defined in Eq.~\ref{eq:amplitude}, and we verified the gauge invariance of the analytical expressions.
The total cross section for each operator $O_i$ is obtained by integrating 
\begin{equation}
\label{eq:linearint}
    \sigma_{\rm NLO}^{(6)}=\sum_i\frac{C_i(\mu)}{\Lambda^2} 2 \Re \left[{\mathcal{M}^{i,(6)}_{\rm NLO}}^{*} \mathcal{M}_{\rm LO}^{(4)} \right]
\end{equation}
over the phase space.
In the following section, we provide numerical results parametrised as 
\begin{equation}
\label{eq:nlo_notation}
    \Delta_{\rm NLO} = \frac{\sigma_{\rm NLO}^{(6)}}{\sigma_{\rm LO}^{(4)}} = \Delta_{\rm NLO}^{\rm finite} + \Bar{\Delta}_{\rm NLO} \log\left(\frac{\mu^2}{s}\right)
\end{equation}
where the cross section $\sigma_{\rm NLO}^{(6)}$ as computed in  Eq.~\ref{eq:linearint} is normalised by the tree-level SM cross section $\sigma_{\rm LO}^{(4)}$. We separate the finite contribution $\Delta_{\rm NLO}^{\rm finite}$ from the leading log contributions $\Bar{\Delta}_{\rm NLO}$ originating from the RGE. The renormalisation scale $\mu$ is typically chosen to be the scale of the process, that is $\sqrt{s}$ in this case.
The parametrisation in Eq.~\ref{eq:nlo_notation} is chosen to facilitate the comparison of the finite effects to those that can be captured via RGE-improved predictions. 
With this convention, different choices of $\mu$ allow to evaluate the impact of the finite part at different EFT scales.

\subsection{Numerical results}
The numerical evaluation is obtained with the PDG input parameter values~\cite{ParticleDataGroup:2026aaa}
\begin{alignat}{2}
    \nonumber
    m_Z &= 91.1879 \, \text{GeV} \, , \quad &&  m_W = 80.3625 \, \text{GeV} \, ,\\
    G_{\mu} &= 1.16638 \times 10^{-5} \, \text{GeV}^{-2} \, , \quad && m_H = 125.13 \, \text{GeV} \, , \\
    \nonumber
    m_t &= 172.6 \, \text{GeV} \, , \quad && m_b = 4.7 \, \text{GeV} \, ,
\end{alignat}
while all other fermion masses have been set to zero.
We focus on the energy runs above the $W^+ W^-$ production threshold at future circular $e^+e^-$ colliders. 
In particular, the FCC-ee would operate at $\sqrt{s}=240$ GeV with an integrated luminosity of 10.8 ${\rm ab}^{-1}$, before reaching the $\sqrt{s}=365$ GeV required to access $t \bar t $ production \cite{FCC:2025lpp}. On the other hand, the LEP3 program foresees a maximum centre of mass energy of $\sqrt{s}=230$ GeV, where it would collect 2.3 ${\rm ab}^{-1}$ of data \cite{Anastopoulos:2025jyh}.
In Tab.~\ref{tab:numerical_results} we provide the NLO numerical result for the two presented running scenarios, where the sensitivity to top-quark operators entirely comes from virtual corrections. Following the notation of Eq.~\ref{eq:nlo_notation}, we present the new results for $W^+ W^-$ production, as well as those computed for $ZH$ production for comparison.
We compared our $ZH$ results to those of~\cite{Asteriadis:2024xts}, finding good agreement.
\footnote{We agree with the results provided in Table 6 of~\cite{Asteriadis:2024xts} except for the finite part of $\mathcal{O}_{uW}^{[33]}$ and $\mathcal{O}_{uB}^{[33]}$. However, the authors also provide their results in their repository \cite{Asteriadis:2024eehz} which was updated since the publication. We find perfect agreement with the latest version of their repository but sizeable differences in the finite part of the dipoles with respect to the numbers quoted in their published paper.} 
The NLO effect of top-quark operators to diboson processes has also been computed in~\cite{Vryonidou:2018eyv,Durieux:2018ggn} using a modified renormalisation scheme and a different reading point prescription for the ambiguous $\gamma_5$ traces. However, we validated our framework by reproducing the results of the Higgs decays provided in~\cite{Vryonidou:2018eyv} in the $\overline{\mathrm{MS}}$ scheme.
\begin{table}[t]
    \centering
    \begin{tabular}{|c|cc|cc||cc|cc|}
    \cline{2-9}
      \multicolumn{1}{c|}{} & \multicolumn{4}{c||}{$e^+e^-\rightarrow W^+W^-$} & \multicolumn{4}{c|}{$e^+e^-\rightarrow ZH$} \\ \cline{2-9}
      \multicolumn{1}{c|}{} & \multicolumn{2}{c|}{$\Delta_{\rm NLO}^{\rm finite}\times 10^{6}$} & \multicolumn{2}{c||}{$\Bar{\Delta}_{\rm NLO}\times 10^{6}$}  & \multicolumn{2}{c|}{$\Delta_{\rm NLO}^{\rm finite}\times 10^{6}$} & \multicolumn{2}{c|}{$\Bar{\Delta}_{\rm NLO}\times 10^{6}$}  \\
    \hline
        $\sqrt{s}$ [GeV]                   & 230    & 240   & 230   & 240   & 230   & 240   & 230   & 240\\ \hline\hline
    $\mathcal{O}_{\varphi q}^{(1)\,[33]}$  & 331    & 359   & 364   & 362   & -872  & -1296 & -3930 & -3942 \\
    $\mathcal{O}_{\varphi q}^{(3)\,[33]}$  & 218    & 203   & -50   & -58   & 174   & 515   &  4192 &  3990 \\
    $\mathcal{O}_{\varphi u}^{[33]}$        & -263   & -302  & -393  & -395  & -926  & -529  &  3375 &  3351 \\
    $\mathcal{O}_{\varphi ud}^{[33]}$       & -18.5  & -19.6 & -10.2 & -10.2 & -125  & -132  &   -83 &   -83 \\
    $\mathcal{O}_{uW}^{[33]}$              & 17.5   &  5.2  & -46.2 & -53.0 & -1503 & -1169 &  3640 &  3873 \\
    $\mathcal{O}_{uB}^{[33]}$              & -100   & -118  & -52    & -59    & 166   & -32   & -2042 & -2145 \\
    $\mathcal{O}_{u\varphi}^{[33]}$        &    0   &    0  &  0    &  0    & 360   & 310   &     0 &     0 \\
    \hline
    \end{tabular}
    \caption{Numerical results for LEP3 and FCCee. The parametrisation is given in Eq.~\ref{eq:nlo_notation}. These values are obtained setting $C_i/\Lambda^2= 1\, \mathrm{ TeV}^{-2}$. The results are normalised by the LO SM cross sections: $\sigma^4_{LO}$=17.815 (17.145) pb at $\sqrt{s}=230 \, (240)$ GeV for $W^+ W^-$ and $\sigma^4_{LO}$=219.6 (239.3) fb at $\sqrt{s}=230 \, (240)$ GeV for $ZH$. }
    \label{tab:numerical_results}
\end{table}
As it appears in Tab.~\ref{tab:numerical_results} we find that NLO corrections are generally smaller for $W^+ W^-$ compared to $ZH$, for both the finite and the logarithmic part. 
In particular, in the case of $ZH$, $\Bar{\Delta}_{\rm NLO}$ is generally much bigger than the finite part $\Delta_{\rm NLO}^{\rm finite}$, by up to more than 50 times in the case of $\mathcal{O}_{uB}^{[33]}$ at $\sqrt{s}= 230$ GeV. As a consequence, any bounds extracted from this process are largely dependent on the scale chosen for the $\overline{\mathrm{MS}}$ renormalisation. 
Conversely, in the case of $W^+ W^-$ the $\Delta_{\rm NLO}^{\rm finite}$ and $\Bar{\Delta}_{\rm NLO}$ contributions are of comparable size for all operators considered. This indicates that purely RGE-induced effects might not, in general, provide a reliable approximation of the full NLO corrections in this process. In particular, finite contributions to diboson production can be phenomenologically relevant, even in global analyses, for new physics scales as high as 1 TeV.
We note here that the top-quark Yukawa effective operator $\mathcal{O}_{u\varphi}^{[33]}$ can only be accessed via $ZH$. Since it does not mix with tree-level operators, it does not have a logarithmic dependence and the sensitivity comes entirely from the finite part.

\begin{figure}[t!]
    \centering
    \includegraphics[width=0.48\linewidth]{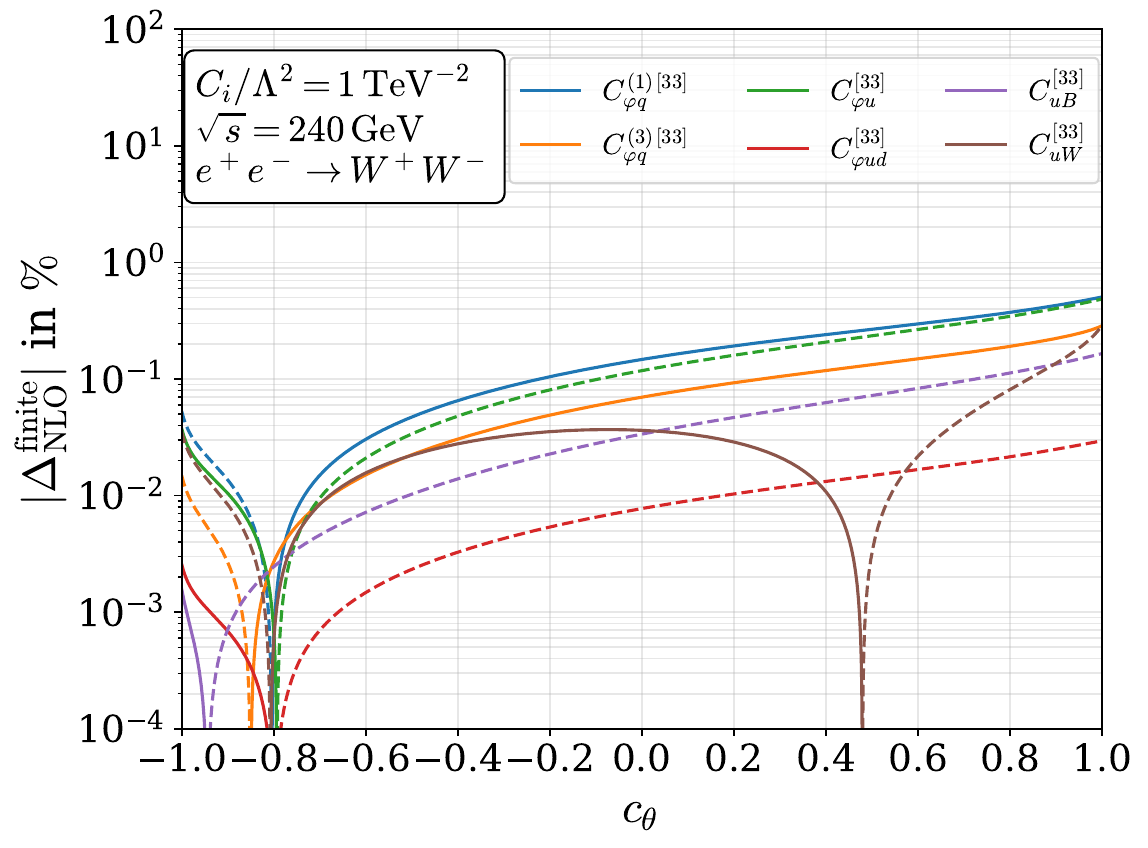}\hfill
    \includegraphics[width=0.48\linewidth]{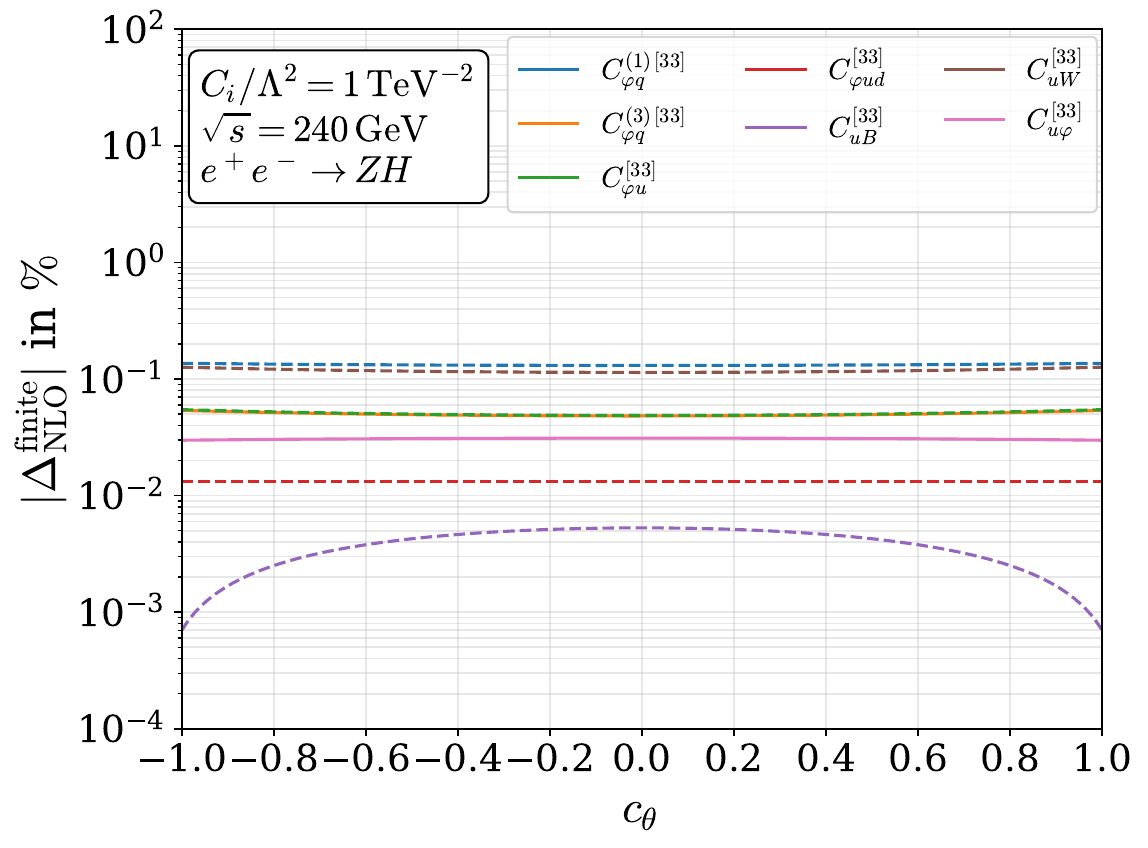}
    
    \caption{The absolute value (in \%) of the differential ratio between the SMEFT and the SM, $\Delta_{\rm NLO}^{\rm finite}$, for \(WW\) (left) and \(ZH\) (right). The dashed lines correspond to negative values of the ratio. Here \(c_\theta=\cos\theta\), where \(\theta\) is the angle between the incoming electron and outgoing \(W^+\) and $Z$ momenta, respectively. 
    }
    \label{fig:diff_xsec}
\end{figure}
%

\begin{figure}[h!]
    \centering
    \includegraphics[width=0.48\linewidth]{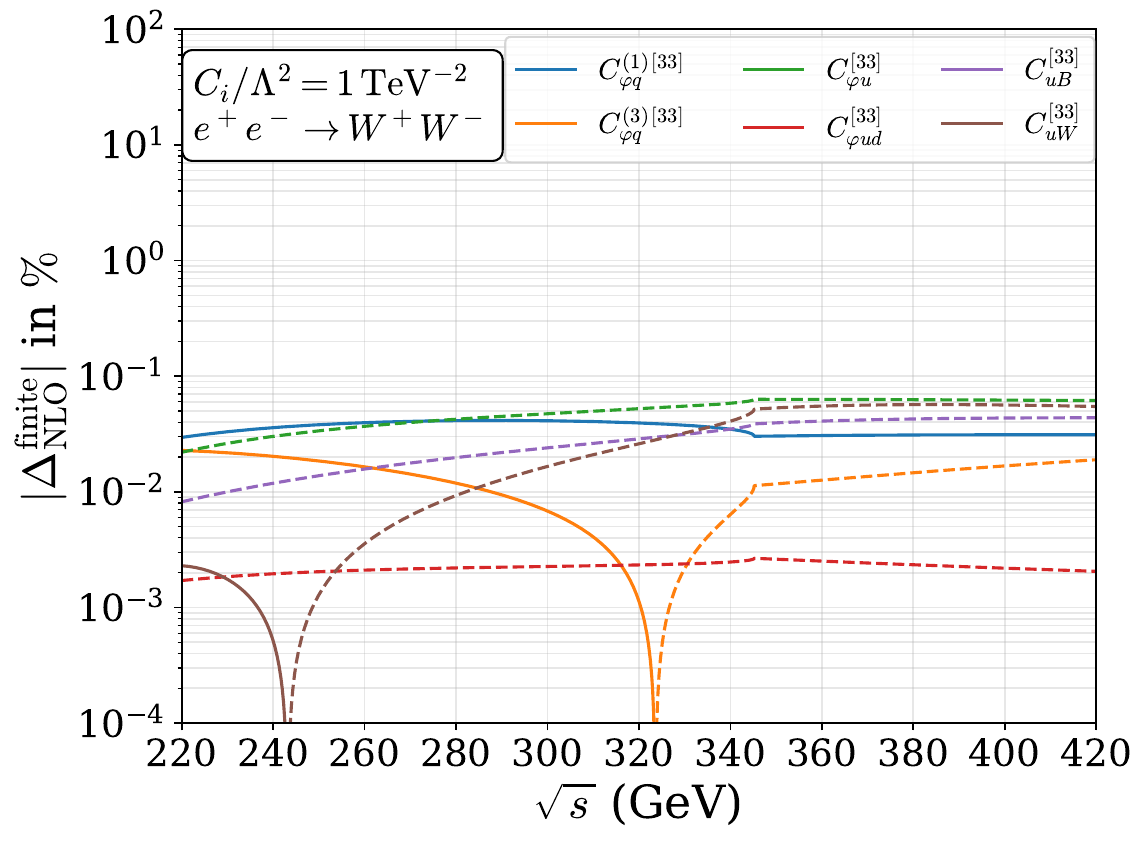}\hfill
    \includegraphics[width=0.48\linewidth]{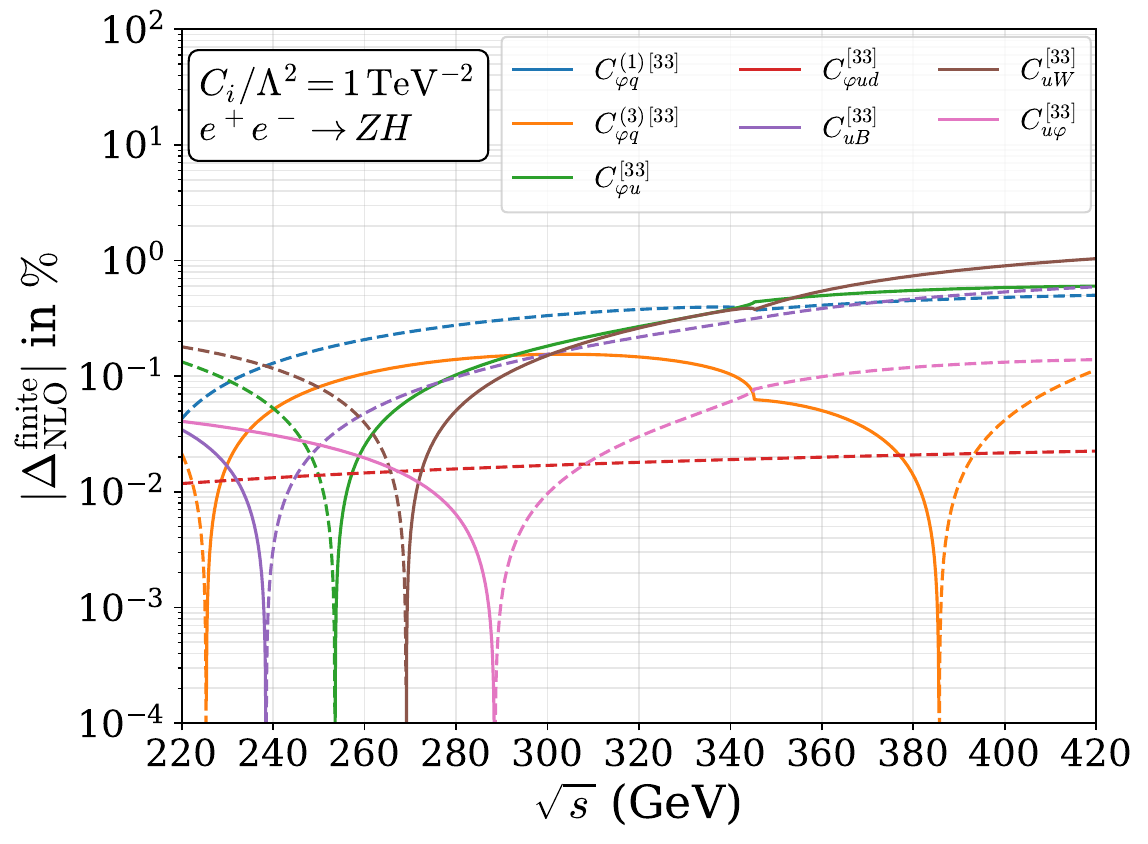}
    
    \caption{The absolute value (in \%) of the inclusive ratio between the SMEFT and the SM, $\Delta_{\rm NLO}^{\rm finite}$, for \(WW\) (left) and \(ZH\) (right). The dashed lines correspond to negative values of the ratio. 
    }
    \label{fig:int_xsec_mu_at_s}
\end{figure}
%

\begin{figure}[h!]
    \centering
    \includegraphics[width=0.48\linewidth]{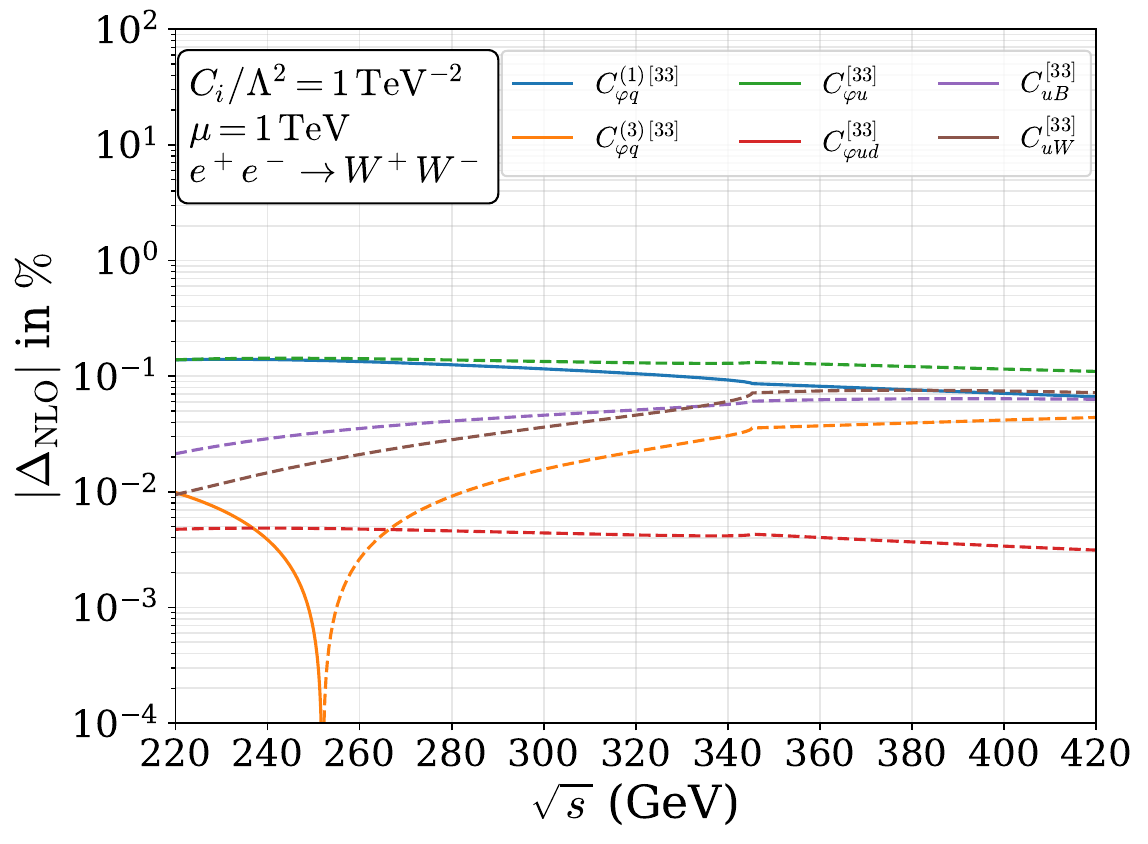}\hfill
    \includegraphics[width=0.48\linewidth]{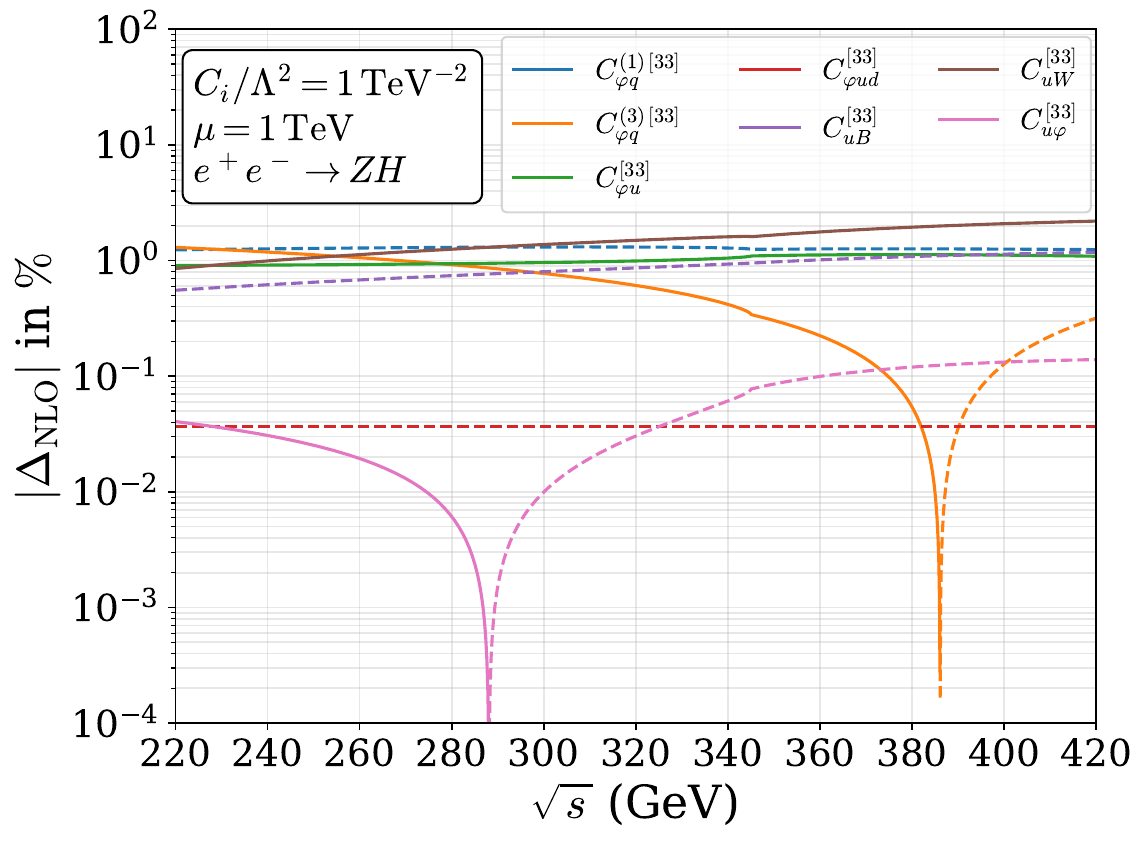}
    
    \caption{
    The absolute value (in \%) of the inclusive ratio between the SMEFT and the SM, $\Delta_{\rm NLO}$, for \(WW\) (left) and \(ZH\) (right) setting $\mu=1$ TeV. The dashed lines correspond to negative values of the ratio. 
    }
    \label{fig:int_xsec_mu_1000}
\end{figure}

The plots in Fig.~\ref{fig:diff_xsec} show the ratio of the differential cross section as a function of the angle of the incoming electron and the outgoing gauge bosons, $\Delta_{\mathrm{NLO}}$, for the $W^+W^-$ (left) and $ZH$ (right) final states, at $\sqrt{s}=240$ GeV. The renormalisation scale is set to the same value, $\mu = 240 $ GeV.  
In the case of $W^+W^-$ production, the differential distributions exhibit a strong angular dependence, and the relative impact of the different operators varies significantly across angular bins. As a result, differential measurements can provide additional discriminating power beyond the inclusive cross section. 
For $ZH$ production, the angular dependence is milder and symmetrical in $\cos\theta$, with the different angular configurations showing similar sensitivities to most operators. This suggests that the inclusive cross-section already captures a large fraction of the available information. 
In Fig.~\ref{fig:int_xsec_mu_at_s} we show for both processes the inclusive cross-section $\Delta_{\mathrm{NLO}}$ as a function of the centre of mass energy, for $\mu=\sqrt{s}$, which corresponds to the finite contribution $\Delta_{\mathrm{NLO}}^{\rm finite}$.
In both processes, we notice a sharp peak in the distribution shape at $\sqrt{s}=345$ GeV, corresponding to the $t \bar t$ on-shell production in triangle diagrams. 
In the case of $ZH$, for all operators the cross-section rapidly changes sign at energies between 220 and 290 GeV, in agreement with the large energy dependence also observed in Tab.~\ref{tab:numerical_results}. On the other hand, for $W^+W^-$ the energy dependence is weaker. 
Finally, in Fig.~\ref{fig:int_xsec_mu_1000} we show the integrated cross-section $\Delta_{\mathrm{NLO}}$ as a function of $\sqrt{s}$, this time for a fixed scale $\mu=1$ TeV. This illustrates the expected sensitivity to each operator contribution at 1 TeV.
In $ZH$, the contribution associated with
$\mathcal O_{\varphi ud}^{[3,3]}$ is generated by electroweak
input-parameter counterterms in the $\alpha_\mu$ scheme, rather
than by an independent correction to the 
$e^+e^-\to ZH$ amplitude. The operator contributes to the $W$-boson
self-energy, which enters the relation between the measured input parameters and the renormalised electroweak couplings. The induced shifts of the tree-level couplings multiply the same structures as
the SM tree-level amplitude.  Consequently, this contribution is
proportional to the SM result and follows the same energy dependence.

\subsection{Phenomenology study}
\label{sec:pheno}
\begin{figure}[t!]
    \centering
    \includegraphics[width=1.\linewidth]{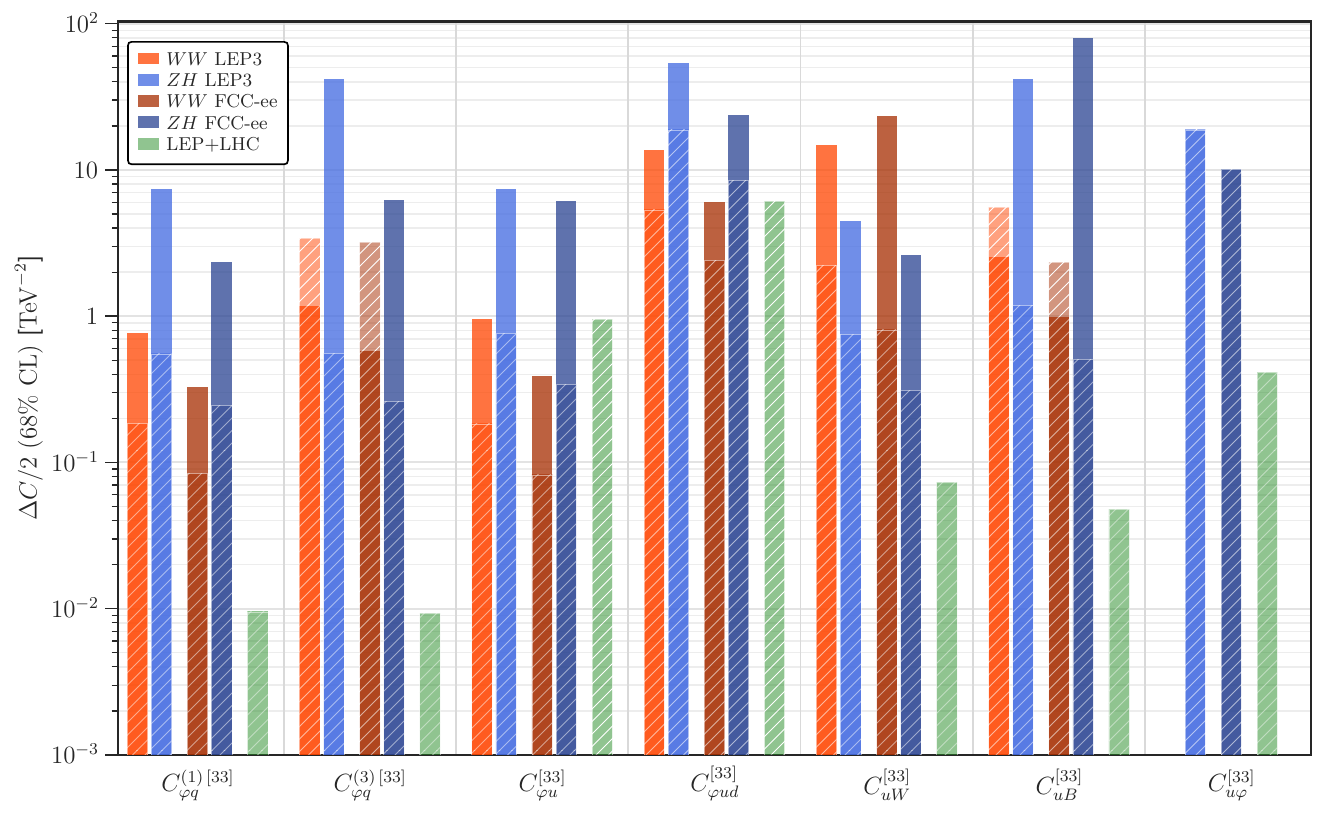}
    \caption{Summary of individual bounds on the Wilson coefficients from several observables. Solid colors correspond to $\mu= 240$ GeV, while the hatched patch corresponds to $\mu=1$ TeV. The individual bounds from the global fit in~\cite{deBlas:2025xhe}, without RGE, are also shown for comparison.}
    \label{fig:bounds}
\end{figure}

To assess the phenomenological impact of these corrections, we show in Fig.~\ref{fig:bounds} the individual bounds for the set of top-quark operators considered in this work for $W^+ W^-$ (red) and $ZH$ (blue), where the lighter and darker shades represent the bounds for LEP3 and FCC-ee respectively. The projected relative precision on the measurements is of
\begin{equation*}
\begin{aligned}
    \mathrm{LEP3}~\mbox{\cite{Anastopoulos:2025jyh}}:\quad & \delta \sigma_{ZH} = 0.67\%, \quad \delta \sigma_{W^+W^-} = 0.0253\% \, , \\
    \mathrm{FCC}\mbox{-}\mathrm{ee}~\mbox{\cite{Selvaggi:2025kmd,DeBlas:2019qco}}: \quad & \delta \sigma_{ZH} = 0.31\%, \quad \delta \sigma_{W^+W^-} = 0.0117\% \, .
\end{aligned}
\end{equation*}
For illustrative purposes, the bounds from  global linear single parameter fits to LEP and LHC data~\cite{deBlas:2025xhe} are also shown, where the RG running was not taken into account. 
The bounds are presented for two choices of $\mu$, the low energy scale $240$ GeV (solid bars), and a high scale $\Lambda=1$ TeV  (hatched bars).
Defining the Wilson coefficients at $\mu \sim \sqrt{s}$ represents the case where the tree level coefficients have already evolved down to the scale of the process $\sqrt{s}$, reabsorbing the logarithmic RGE contributions. The sensitivity is then typically weaker (depending on the sign and size of $\Bar{\Delta}_{\rm NLO}$) and the bounds are entirely determined by $\Delta_{\rm NLO}^{\rm finite}$. 
On the other hand, assuming a high scale $\mu= \Lambda$ is equivalent to taking into account the single-logarithmic RGE contribution. The bounds are generally tighter in this case due to the logarithmic enhancement. The only exceptions to this behaviour are in $W^+ W^-$ for the coefficients $C_{\varphi q}^{(3)\,[33]}$ and $C_{uB}^{[33]}$. This happens because $\Delta_{\rm NLO}^{\rm finite}$ and $\Bar{\Delta}_{\rm NLO}$ have opposite sign and, for $\mu= 1$ TeV, $|\Delta_{\rm NLO}^{\rm finite}|>|\Delta_{\rm NLO}^{\rm finite}+\Bar{\Delta}_{\rm NLO}\log(\mu^2/s)|$. 
As expected, the scale choice has a significant impact on the results. However, this is typically less pronounced for $W^+ W^-$ than for $ZH$, suggesting that the finite part might be relevant even in the context of a global fit, and that the full dependence on some coefficients cannot be fully captured by the RGE logs.
Many of the bounds from $W^+ W^-$ are better than for $ZH$, also thanks to the higher statistical precision expected which compensates for the relatively smaller NLO effects. 
Comparing with the current bounds from the global fit, we see that some of the constraints are competitive with the current fit results, in particular $C_{\varphi u}^{[33]}$ and $C_{\varphi ud}^{[33]}$. A combination of the two processes as well as information from EWPOs and Higgs decays in a global fit for future colliders will allow to probe top-quark interactions beyond the current precision even in the absence of runs above the $t\bar{t}$ threshold. 

\section{Conclusions}
\label{sec:conclusions}
In this work we computed the NLO EW corrections to $e^+e^-\to W^+W^-$ from dimension six top-quark operators at order $\mathcal{O}(1/\Lambda^{2})$ in the SMEFT, motivated by the high precision program of future electron–positron colliders. 
This work constitutes a first step toward the full computation of NLO corrections. Our results demonstrate that EW corrections to diboson production provide sensitivity to top-quark operators even at centre-of-mass energies below the $t \bar t$ production threshold, yielding constraints competitive with current bounds from LEP and LHC data. %
A systematic study of the full NLO SMEFT dependence, which is left for future work, will be crucial to fully understand the potential of diboson measurements at future colliders. The combination with a wider set of processes will be essential for exploiting global SMEFT analyses and enhancing the sensitivity to new physics.

\acknowledgments
E.C. and V.M. thank Giuseppe Ventura for helpful discussions during the course of this work.
The work of V.M. is supported by the MICIU through a Beatriz Galindo Junior grant (BG24/00038), and by MICIU/AEI/\allowbreak10.13039/\allowbreak501100011033 through grant PID2025\allowbreak-171\allowbreak322NB\allowbreak-C21. E.C. and E.V. are supported by the European Research Council (ERC) under the European Union’s Horizon 2020 research and innovation programme (Grant agreement No. 949451) and a Royal Society University Research Fellowship through grant URF/R1/201553.

\appendix

\bibliographystyle{JHEP}
\bibliography{biblio}

\end{document}